# Spiking Neural Network Based Low-Power Radioisotope Identification using FPGA

Xiaoyu Huang[1], Edward Jones[2], Siru Zhang[3], Shouyu Xie[1], Steve Furber[2], *Fellow, IEEE*, Yannis Goulermas[3], Edward Marsden[4], Ian Baistow[4], Srinjoy Mitra[1], Alister Hamilton[1]
[1]University of Edinburgh, UK, [2]University of Manchester, UK, [3]University of Liverpool, UK, [4]Kromek Group plc, UK
E-mail: xiaoyu.huang@ed.ac.uk, alister.hamilton@ed.ac.uk

***Abstract*—this paper presents detailed methodology of a Spiking Neural Network (SNN) based low-power design for radioisotope identification. A low power cost of 72 *mW* has been achieved on FPGA with the inference accuracy of 100% at 10 *cm* test distance and 97% at 25 *cm*. The design verification and chip validation methods are presented. It also discusses SNN simulation on SpiNNaker for rapid prototyping and various considerations specific to the application such as test distance, integration time and SNN hyperparameter selections.**

## I. INTRODUCTION

The threat of nuclear weapons and radiological dispersion devices (dirty bombs) is ever present and for this reason, technologies for the detection and identification (ID) of radioisotopes play an essential role in national security. Previous work has seen the application of a number of algorithms and machine learning techniques to the task of automated radioisotope ID [1]. Most of this work has concentrated on the accuracy of ID and the proposed approaches almost exclusively use histogram-type data, built up from the integration of incoming detection events in discrete frames. Power consumption is rarely considered or analysed, something that is an important consideration when such algorithms are applied in edge devices. In [2], the problem of high power overhead for frame-based methods was identified and an alternative event-based processing method proposed but without detailing hardware implementation or results. The motivation of the work in this paper is to establish a detailed and comprehensive hardware design methodology for low power event-based radioisotope identification.

### *A. The Problem*

The conventional radioisotope ID process is typically implemented in a frame-based manner. The problem with this approach from an energy perspective is that the data processing units continuously consume power, even when the event arrival rates are very low or zero. This is due to the lack of event-aware processing.

One example of a frame-based approach is shown in Fig. 1 (a). The integration time is normally a couple of seconds depending on the sensitivity of the photodetector. In this example two seconds is chosen for simplicity. As demonstrated in Fig. 1 (b), the whole process for each frame takes two seconds of Data Sampling and one second of Data Processing in series, and the one second inference can be achieved through pipelining. Due to the lack of event-aware processing, the process units continuously consume power even when only being in the presence of background radiation. Current solutions available on the market have a processing overhead for the radioisotope ID algorithm that requires the digital signal to be broadcast to a mobile phone via Bluetooth and processed using an app. This is necessary to keep the detector's battery life, size and weight down. The power consumption for the whole procedure is reported to be around 2 *W*.

### *B. Event-based Processing*

The problem established above can be addressed by using an event-based processing method. This means that processing is initiated only when detection events take place, giving the potential for improved energy efficiency. Fig. 1(b) illustrates the event-based radioisotope ID process where the processing is carried out in parallel with the events only in an asynchronous manner and an early stopping inference is likely possible for strong sources.

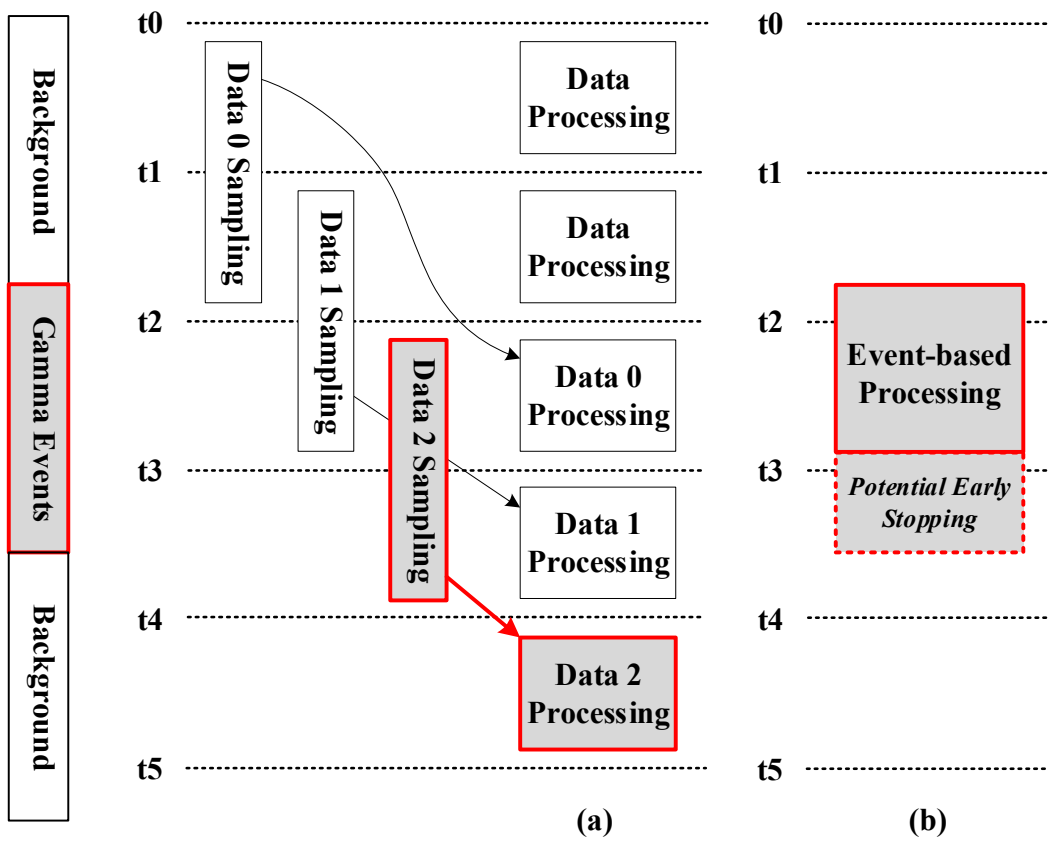

Fig. 1 (a) Frame-based and (b) event-based signal processing

In this paper, the event-based processing is realised by Spiking Neural Networks (SNNs) that are the output of conversion from artificial neural networks (ANNs). These SNNs are simulated and evaluated on the SpiNNaker [3] neuromorphic hardware platform first to speed up the prototyping process. They are then implemented on a Field-Programmable Gate Array (FPGA) for a proof-of-principle. To our knowledge, this is the first hardware implementation of SNN based radioisotope ID achieving extremely low power consumption. Section II-V presents the system level architecture and the methodology of the hardware implementation of the spiking neural network (SNN) processor. Section VI describes the experiments undertaken to evaluate this implementation, using the capabilities of the SpiNNaker neuromorphic platform to evaluate the algorithm before implementation on an FPGA. Finally Section VII discusses the findings and outlines future work.

## II. SYSTEM ARCHITECTURE

Fig. 2 outlines the system-level components of our proposed radioisotope ID system. The process is divided into the detector steps, data pre-processing steps, and our proposed event-based processing step. The rest of this section will describe these steps.

### A. *Sensor*

The invisible gamma photons are converted into visible photons by the scintillator, and then the visible photons are captured by the Photon Detector (PD).

### B. *Data Pre-processing*

*1) Analogue-to-event Conversion:* For each gamma event the analogue signal from the PD, which represents the energy level of the incident photon, is converted into a digital event of a particular energy channel bin by the Analogue to Event Converter (AEC), based on the multiple-threshold technique in [4].

*2) Dimensionality Reduction:* The event data are then pre-processed by undergoing dimensionality reduction. Our preliminary experiments indicated that simple re-binning was sufficient to reduce the input complexity for the hardware design and potentially save power. Rebinning is a simple method that serves to uniformly downsample the energy spectrum by combining adjacent energy bins.

### C. *Event-based Processor*

The pre-processed event data is fed into the event-based processor, which is a hardware implementation of an SNN, for classification. The SNN is shown in the Fig. 3.

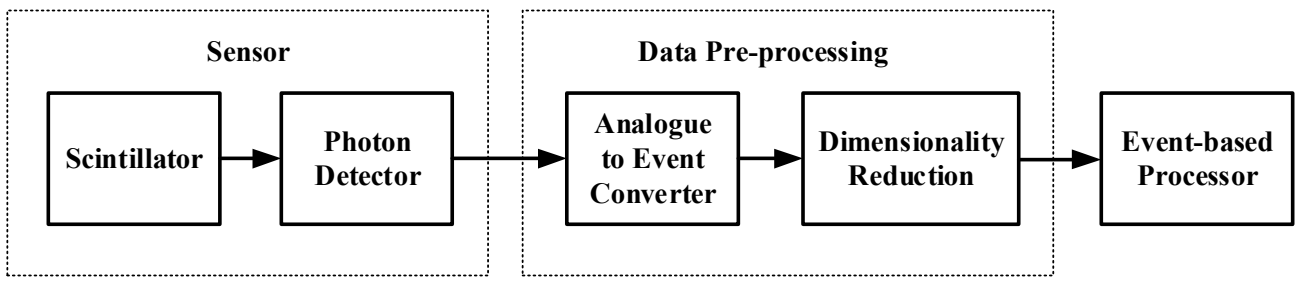

Fig. 2 System level architecture

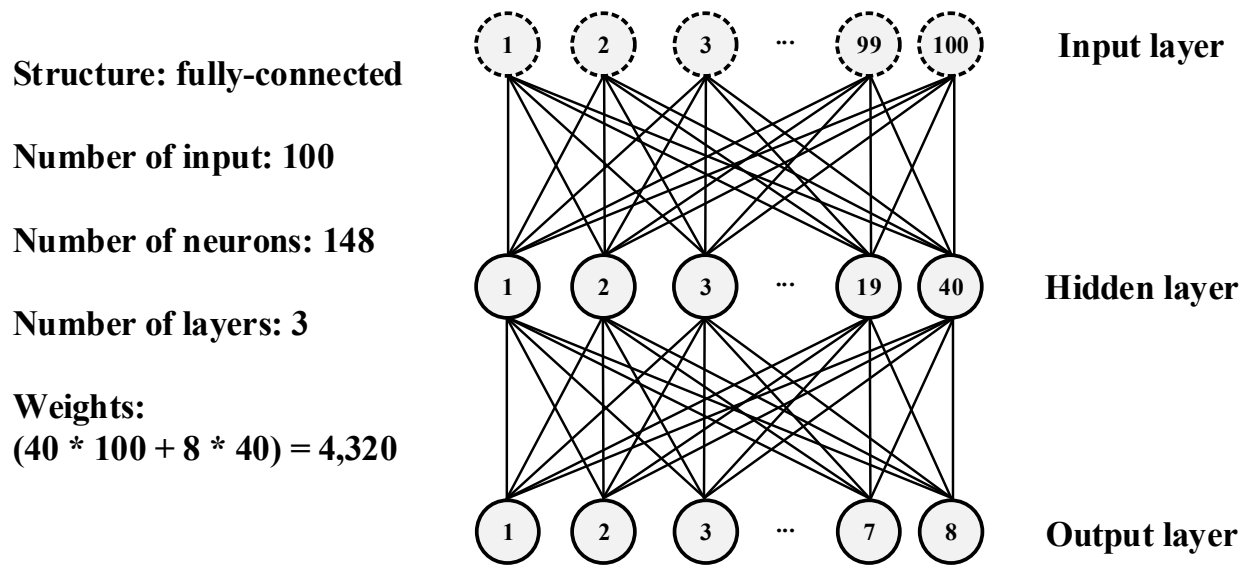

Fig. 3 SNN architecture

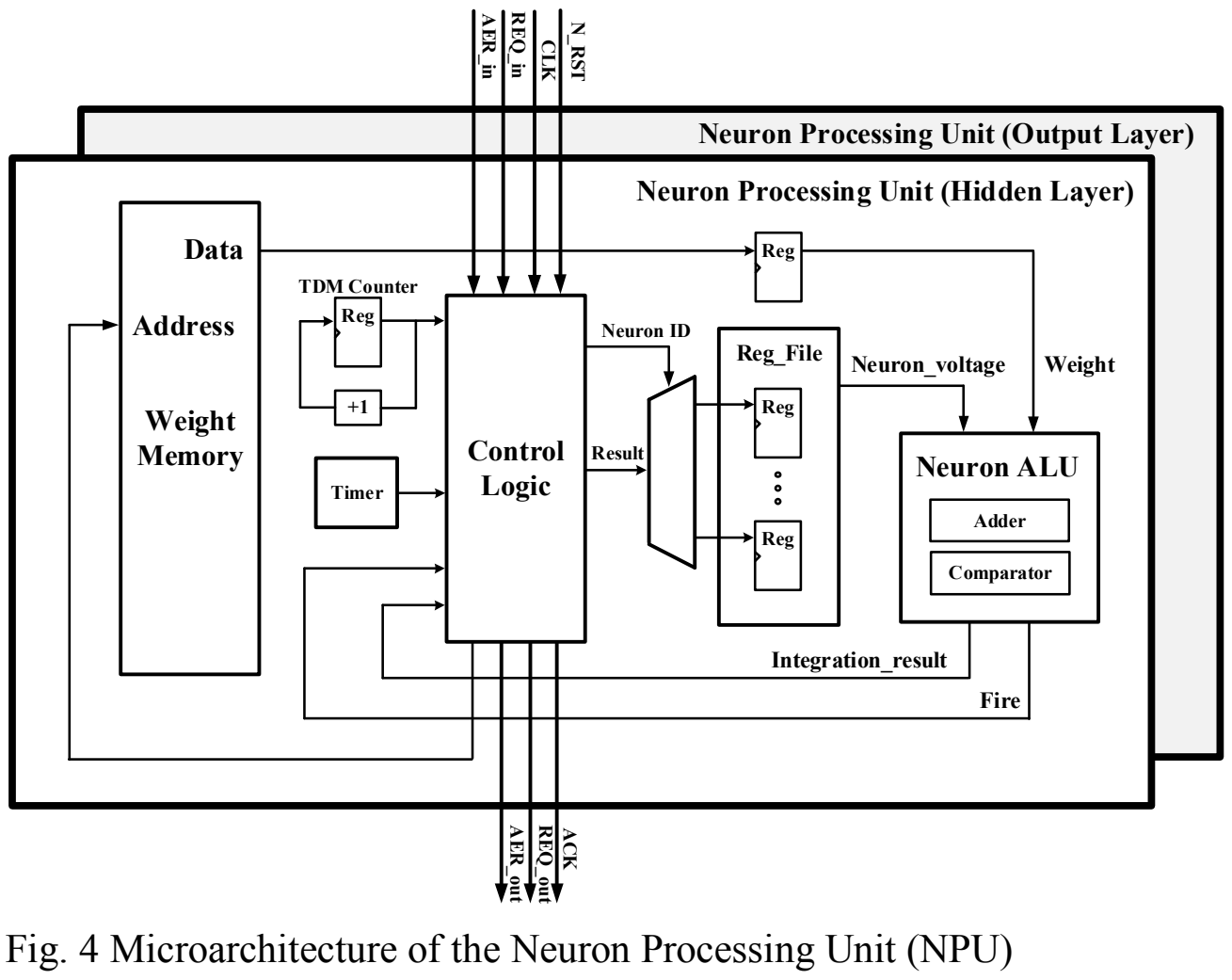

Fig. 4 Microarchitecture of the Neuron Processing Unit (NPU)

## III. NEURON PROCESSING UNIT

Each layer of the SNN is implemented in the microarchitecture shown in Fig. 4. The Address Event Representation (AER) protocol [5] is used for the communication between each layer, and a memory is shared and used for the storage of all the synaptic weights.

Each neuron in the same layer triggered by the same pre-synaptic neuron, is processed one-by-one in a Time Division Multiplexing (TDM) manner. The sequence of the operations and data flow are managed by the state machine in the Control Logic block.

As shown in Fig. 5, to handle the worst case of back-to-back spikes from the hidden layer, 9 cycles of delay are added after each spike fire event of any neurons in hidden layer to allow sufficient time for the 9 cycles needed by TDM process of 8 neurons in the output layer.

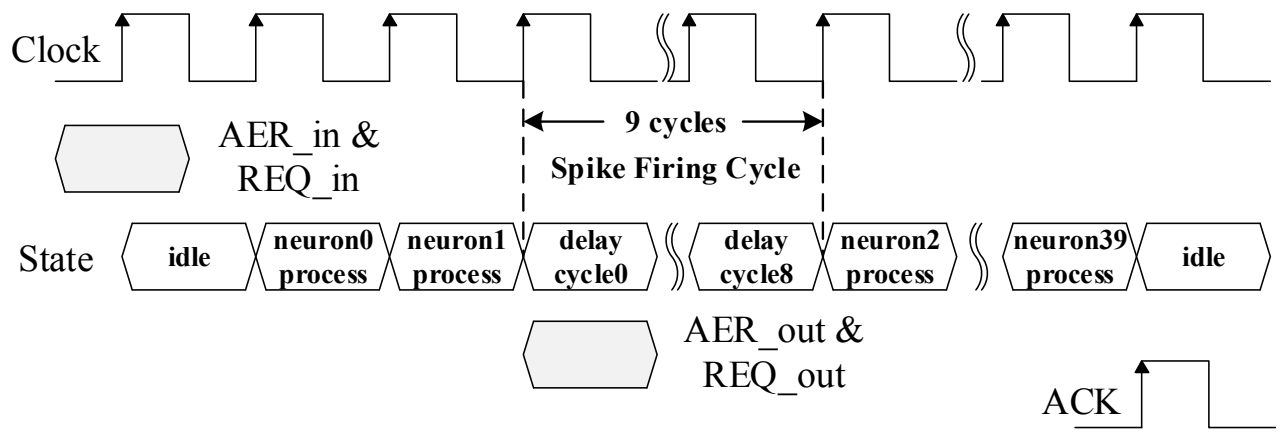

Fig. 5 Timing diagram of one transaction example in hidden layer

## IV. NEURON MODEL

The neuron model used in this work is based on the Integrate-and-Fire (IF) model. To reduce power and cost, all unnecessary features such as the refractory period, alpha-function-shaped synaptic current and membrane voltage leak are not included. A SNN with the same simplifications in [6] demonstrated high inference performance.

$$V(t) = V(t-1) + \textstyle\sum_i x_i\,(t-1)w_i + L, \qquad (1)$$

$$V(t) = V(t-1) + w_i, \qquad (2)$$

$$if\ V(t) \geq V_{thr},\quad Fire\ and\ Reset\ V(t) = 0, \qquad (3)$$

The dynamics of the original IF model are described by equation (1). The membrane potential $V$ of the neuron is the sum of previous state, all the active synaptic inputs weights and a leak parameter $L$. The $x_i(t-1)$ represents either the presence (1) or absence (0) of a spike from the pre-synaptic neuron. The leak parameter $L$ is zero in this work. In this application, equation (1) can be further simplified into equation (2) based on the fact that only one neuron can fire in each layer during single time step (1 $\mu s$) due to the sparsity of the gamma decay event (less than 1 kHz Poisson rate) and the limitation of the photon detector.

The operations in the equations above can be implemented by the Neuron Arithmetic Logic Unit (ALU) shown in Fig. 4. Only an adder and a comparator are needed. The adder is used for the integration in equation (2) and the comparator is used for equation (3).

## V. VERIFICATION AND VALIDATION

Due to the potential complexity of SNNs, especially as network size increases, the performance test is not enough to ensure the quality of the design and provides very little information for debugging. Multiple failures of the extremely long performance tests in simulation or bugs found late in the validation stage will increase the design cycle dramatically and delay the time-to-market. For these reasons, design verification and chip validation consideration at the early stage are essential.

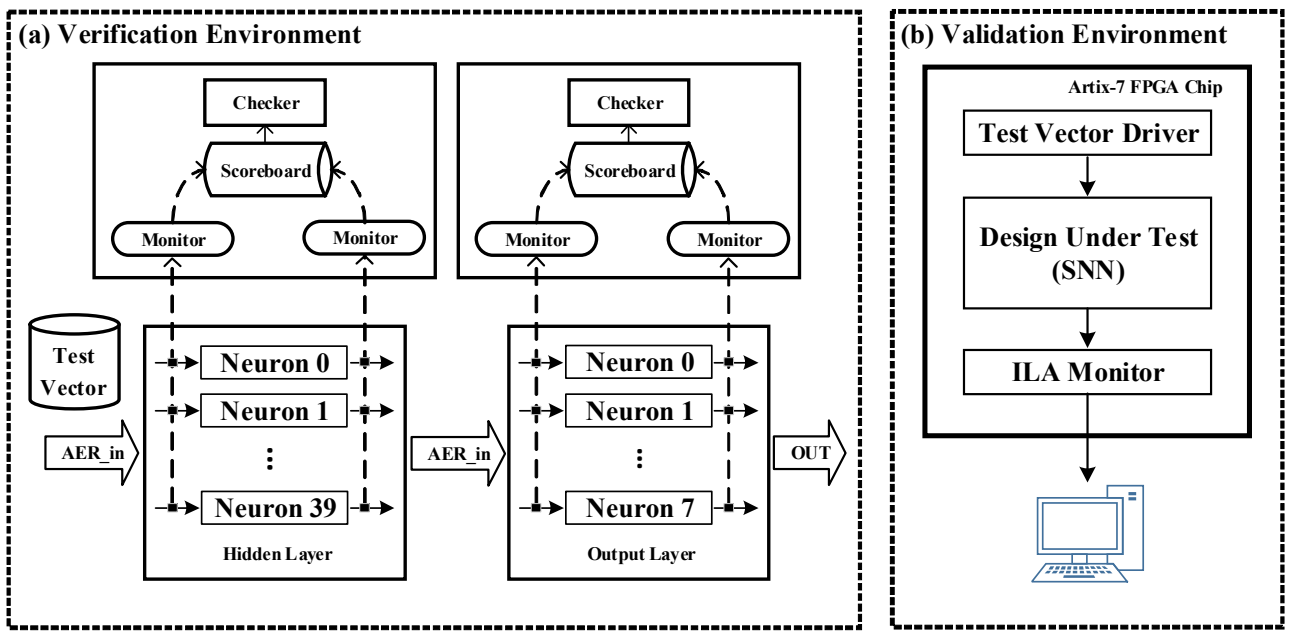

Fig. 6 (a) Design verification structure; (b) Chip validation structure

Fig. 6 shows the verification and validation structures. As shown in Fig.6 (a), monitors, scoreboards and checkers are inserted for each neuron in the SNN to check that the status of each neuron is as expected. Mismatch in the neuron membrane voltage is reported during the simulation which allows for faster diagnosis of a failed test.

Once the SNN is programmed on FPGA, the functionality and the performance of the chip can be tested using the validation structure shown in Fig. 6 (b). A test vector driver is coded in Verilog and programmed on the same FPGA to provide stimulus, and a Vivado built-in IP of Integrated Logic Analyser (ILA) is integrated into the validation system to monitor the inference results delivered by the SNN.

## VI. Experiments and Results

### A. *Dataset and the ANN Training*

The dataset is collected from the experiment with the same setup in [2]. The raw dataset is made up of eight industrial radioisotopes ($^{241}$Am, $^{133}$Ba, $^{57}$Co, $^{60}$Co, $^{137}$Cs, $^{152}$Eu, $^{226}$Ra, $^{232}$Th) and was recorded using two different measurement setups, with and without a Polymethyl Methacrylate (PMMA) phantom, designed to model the upper torso of a user. Spectra for each radioisotope source under test were recorded every second for 120 seconds. Subsequent measurements were made at varying distances (10 *cm*, 25 *cm*, 50 *cm*, 1 *m* and 1.5 *m)* with each source. The size of dataset obtained from the tests is therefore 9,600.

The ANN is trained based on 70% of the dataset which covers all the scenarios with even distribution. To study the impact of different distances, the ANN was given groups of examples of the same distance to test the performance of the classifier.

Based on the experiments, it was found that the accuracy of the classifier was affected by the following two factors:

*1)* Data Quality: the data quality deteriorates when the distance is above 10 cm due to the intrinsic limitation of the detector and small size of the dataset. The accuracy of the method drops when the distance is increased as illustrated in Fig. 8 (left).

*2)* *Radioactivity:* From the dataset visualisation shown in Fig. 7, the strengths of data for different radioisotope are different due to their inherently different levels of radioactivity. The algorithm shows the least accuracy when processing the signal of Th-232. With 1-second integration time, it is indistinguishable with background noise.

Based on the whole dataset, Fig. 8 (right) presents the ANN accuracy of test results with different neuron numbers in the hidden layer. Neuron number of 40 shows the highest accuracy, which is taken for the hardware implementation. Figure 7 (left) shows the re-binned gamma-ray spectra of samples of each isotope, this shows how the different isotopes have different signatures which can be identified but also have significant overlaps.

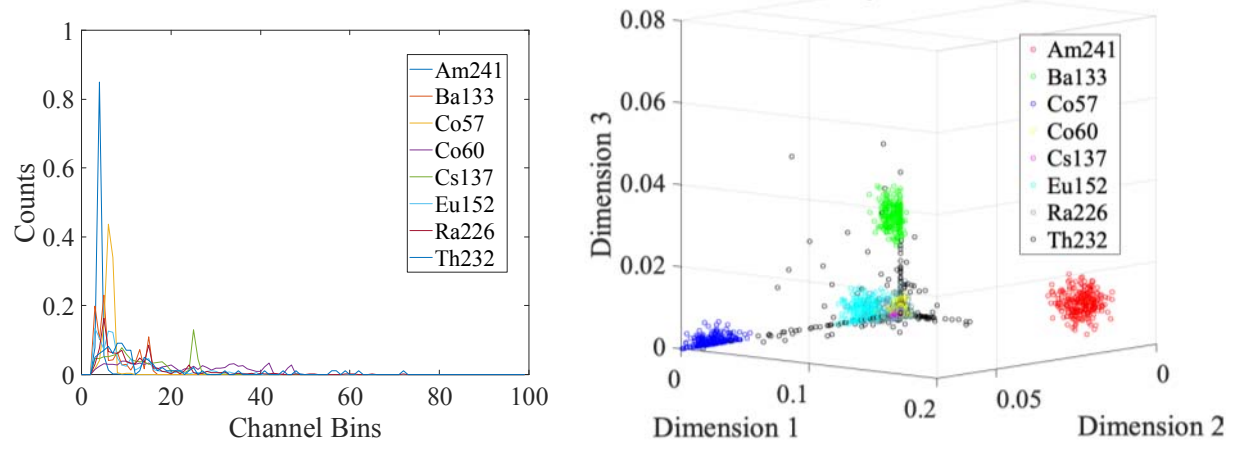

Fig. 7 Data visualisation of the normalised dataset for all 8 isotopes under the test distance of 10 *cm*: (left) 2-D re-binned histogram; (right) 3D representation of the dataset based on data pre-processed using non-negative matrix factorisation.

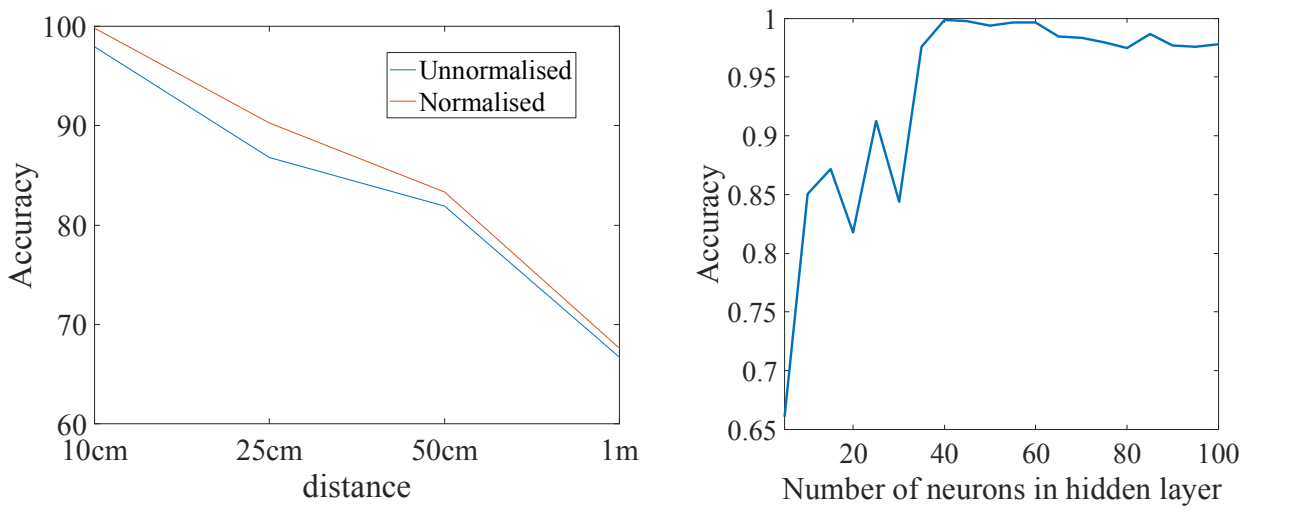

Fig. 8 Accuracy of the ANN with different test distances (left) and different neuron numbers in the hidden layer (right)

### B. *ANN-to-SNN Conversion*

The process of conversion from the trained ANN to a SNN model running on SpiNNaker was made up of two steps: quantisation and network conversion. In the quantisation step the ANN parameters were fine-tuned using quantisation aware training available in the TensorFlow Model Optimization Toolkit [9] and the weights were quantised to 8-bit signed integers. Figure 9 (right) shows the difference that the losses due to weight quantisation were negligible. This suggests that the model has a high number of redundant connections and that reduction of the model complexity through further reduction of the precision of the weights or through pruning of connections could give further energy savings in the hardware implementation. Lower precisions were not investigated here due to the lack of software support for precisions lower than 8-bit integer.

In the network conversion step, we used a version of the SNN toolbox as developed by Rueckauer et al. [7], modified to allow use SpiNNaker as a backend. The neuron model used is a standard PyNN current-based leaky integrate-and-fire neuron model with delta input (LIF_curr_delta) [8] with the default parameters given in the SNN toolbox. This conversion process maps each ANN neuron to an SNN neuron, each ANN weight to SNN weight and the frame-based input histograms counts to the rates of Poisson sources (event generators). This method sees event data synthesised from the frame-based histograms such that the SNN model sees event-based input, as it would from a detector, but still allows the model to be trained using frame-based data using established tools.

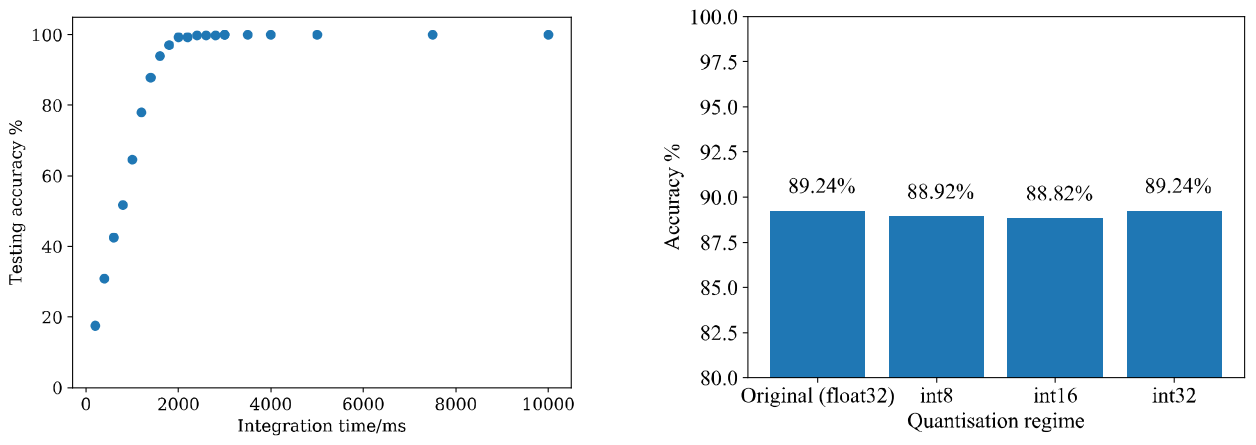

Fig. 9 (left) Integration time curve of the radioisotope identification SNN, based on dataset with distance of 10 cm and a global input spike rate of 500Hz; (right) Comparison of model test accuracy for signed integer quantised ANNs

## C. *SpiNNaker Simulation Results*

SpiNNaker is a neuromorphic hardware platform designed for the simulation of large-scale spiking neural networks at speeds close to biological real-time [3]. In this work SpiNNaker is used to verify model behaviour before implementation using an FPGA. The energy efficiency and scale of SpiNNaker systems allows for rapid prototyping of network architectures. Figure 9 (left) shows how SpiNNaker was used to evaluate the accuracy of a radioisotope identification SNN across a range of integration times. These results were used to establish the 3000 *ms* integration time for model evaluation in trials at a greater distance, with the global input spike rate of 500 Hz this equates to approximately 1500 input spike events on average. In this work we used parallel model execution in real-time to perform a parameter sweep for one parameter. Models could be tuned using more advanced parallel optimisation algorithms such as a genetic algorithms [9].

## D. *FPGA Results*

Hardware implementation results are reported in Table I. The SNN has been implemented on the FPGA with a maximum clock frequency of 100 MHz.

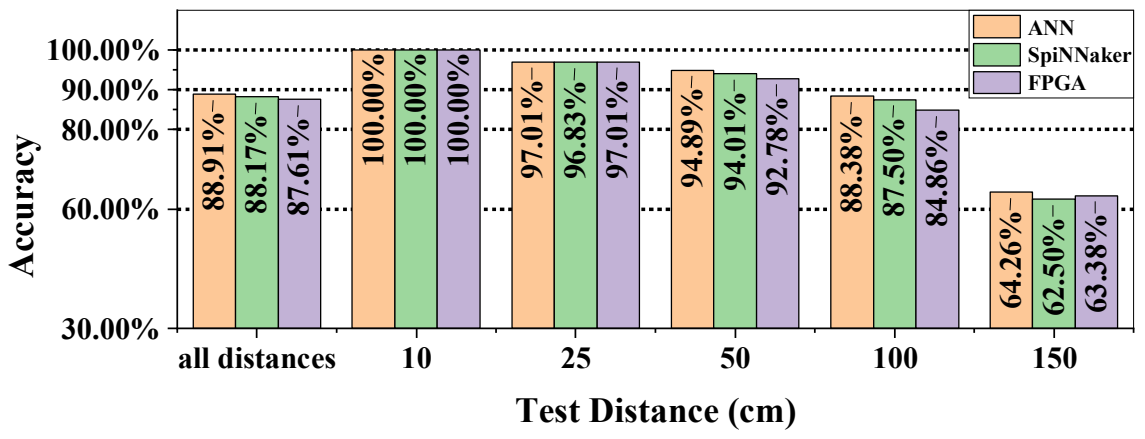

Fig. 10 the inference accuracy results of the ANN, SpiNNaker and FPGA based on a dataset of 2,840 radioisotopes tested in five distances

The average power consumption per inference is 72 $mW$, which is estimated by the Vivado power analysis tool based on the Switching Activity Interchange format (SAIF) file produced from the post-implementation timing simulations. This is more than 27 times better than the 2 $W$ power consumption of the frame-based method. Table II shows the breakdown of the power consumption. The major power cost is from the static power and clocks. The processing power cost is less than 2% of total power consumption, which is due to the low event signal rate of 500 Hz.

A comparison of inference accuracy between ANN, SpiNNaker and FPGA is illustrated in Fig. 10. Those results are based on the testing dataset which covers all the scenarios with even distribution. The FPGA implementation achieved the same level of accuracy in most of the cases with extremely low power and area cost.

TABLE I. LOW-LEVEL DEVICE UTILIZATION OF THE SNN

| Logic Utilisation | Used | Available | Utilisation |
|---|---|---|---|
| Slice Registers | 481 | 41600 | 1.16% |
| Slice LUTs | 986 | 20800 | 4.74% |
| BUFGCTRL | 1 | 32 | 3.13% |

TABLE II. BREAKDOWN OF ON-CHIP POWER CONSUMPTION

| Dynamic Power ($mW$) | | | | Static Power ($mW$) |
|---|---|---|---|---|
| Clocks | Signals | Logic | I/O | |
| 2 | $< 1$ | $< 1$ | $< 1$ | 70 |

# VII. CONCLUSION

A low power event-based hardware design for radioisotope identification has been presented. This paper has detailed the methodology from initial design to FPGA implementation and presented results. For ease of comparison between different isotopes and distances, normalised histograms were used as inputs to models here, future work will need to incorporate gain control into the model to modulate the network inputs and allow the hardware to work with a wider range of input rates.